\documentclass[a4paper]{jpconf}
\usepackage{graphicx}
\usepackage{comment}
\usepackage{subfig}
\usepackage{amssymb,amsmath}
\newcommand{\m}            {\, \mbox{${\rm m}$}}

\newcommand{\snn}          {\ensuremath{\sqrt{s_{\rm NN}}}}

\newcommand{\pT}{\ensuremath{p_{\rm T}}}
\newcommand{\dpT}{\ensuremath{\delta\pT}}

\newcommand{\gevc}{\ensuremath{\mathrm{GeV}/c}}

\newcommand{\kt}          {\mbox{$k_{\rm T}$}}
\newcommand{\antikt} {\mbox{anti-$k_{\rm T}$}}
\renewcommand{\phi}          {\varphi}

\begin{document}
\title{Measurement of jet \pT{} spectra in Pb--Pb collisions at $\snn=2.76~\mathrm{TeV}$ with the ALICE detector at the LHC}

\author{Salvatore Aiola$^{1,2}$}

\address{$^1$ University of Catania and INFN Sezione di Catania,  95123 Catania, Italy}
\address{$^2$ Lawrence Berkeley National Laboratory, Berkeley, CA 94720}

\ead{Salvatore.Aiola@cern.ch}

\begin{abstract}
Reconstruction of jets in high-energy heavy-ion collisions is challenging due to the large and fluctuating
background coming from the underlying event. We report results on full jet reconstruction, 
obtained from data collected in 2011 by the ALICE detector at LHC for \mbox{Pb--Pb} collisions 
at $\snn=2.76$~TeV. The analysis makes use of the tracking system and the electromagnetic calorimeter.
Signal jets, which come from hard scattered partons, are reconstructed using the \antikt{} jet finder algorithm. 
The average background is subtracted on a jet-by-jet basis to 
reduce the contribution to the jet reconstructed energy coming from the underlying event. The jet spectrum 
is corrected to account for fluctuations in the background momentum density and detector effects through 
unfolding.
\end{abstract}

\section{Introduction}
QCD jets are produced in high-energy particle collisions as a result of the fragmentation of a high momentum scattered parton.
Experimentally they are reconstructed using a well-defined algorithm, which acts as a working definition of a jet, to be used
consistently also in phenomenological models.
In heavy-ion collisions, hard 
scattered partons are produced in the early stages of the collision, so that they propagate through 
(and potentially are affected by) the hot and dense nuclear medium. The interactions suffered by 
the parton can result in energy loss, widening and/or complete absorption of jets. These phenomena go under the name
of ``jet quenching''.
RHIC and LHC experiments have already collected convincing evidence of jet quenching in a 
number of measurements, such as the suppression
of high-\pT{} particles relative to a pp baseline~\cite{hadrons-star-03,hadrons-phenix-04,hadrons-alice-10,hadrons-cms-12}, 
hadron-hadron correlations~\cite{dijet-star-06,dijet-alice-12}
and jet-jet correlations at high-\pT{}~\cite{dijet-atlas-10,jet-cms-11}. However, performing
a full jet reconstruction in the low and intermediate \pT{} region is particularly challenging due to the overwhelming soft particle background 
coming from the underlying event (UE). ALICE has recently reported
on a first detailed study of the background for jet reconstruction in the heavy-ion environment~\cite{bkg-alice-12}.

In these proceedings the analysis techniques utilized to fully reconstruct jets in Pb--Pb collisions with the 
ALICE experiment are presented. The results shown here come from data collected by ALICE in fall 2011 at $\snn=2.76$~TeV.

\section{Inputs to the jet finder}
The ALICE tracking system benefits of
the Inner Tracking System, a six-layer silicon detector which provides a precise measurement 
of the primary vertex together with the first points 
of the tracks, and of a large Time Projection Chamber~\cite{alice-08}. 
Tracks are reconstructed at mid-rapidity ($|\eta|<0.9$) and in full azimuth.
The ALICE electromagnetic calorimeter~\cite{ppr-emcal} is a Pb-scintillator sampling calorimeter, which covers mid-rapidity
($|\eta|<0.7$) and partial azimuth ($\Delta\phi=100^{\circ}$). The EMCal measures photons, e.g. from $\pi^0$ decays, which are
included in the jet finder input.
The shift of the jet energy scale due to 
unreconstructed particles, such as $K^0_L$ and neutrons, has been studied in pp simulations and 
accounted for in the final result.
A full description of the ALICE experiment is available in Ref.~\cite{alice-08}.

\section{Jet finding and average background}
\begin{figure}[t]
\centering
\subfloat[][$R=0.2$.]
{\includegraphics[width=.4\columnwidth]{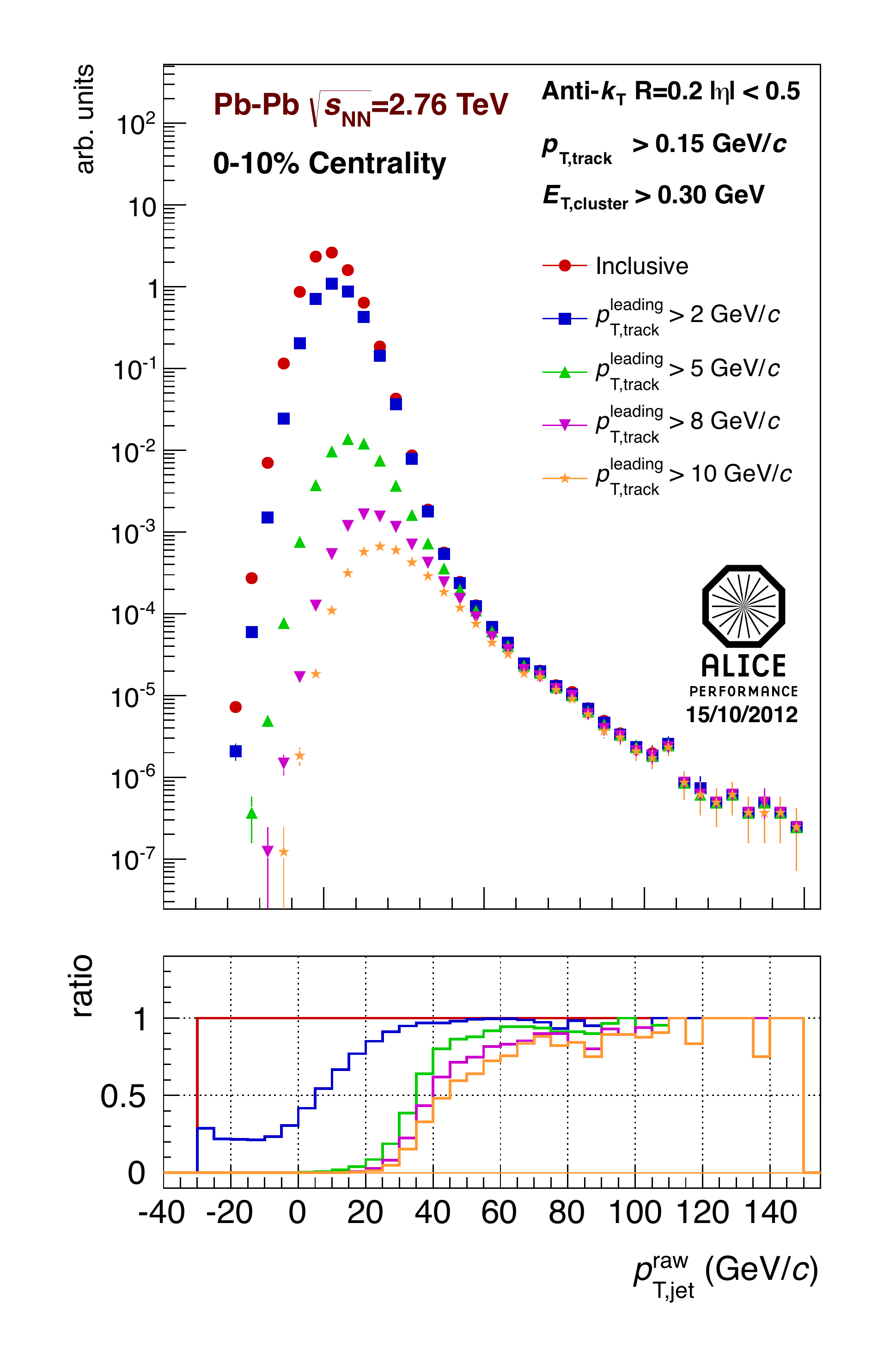}} \quad
\subfloat[][$R=0.3$.]
{\includegraphics[width=.4\columnwidth]{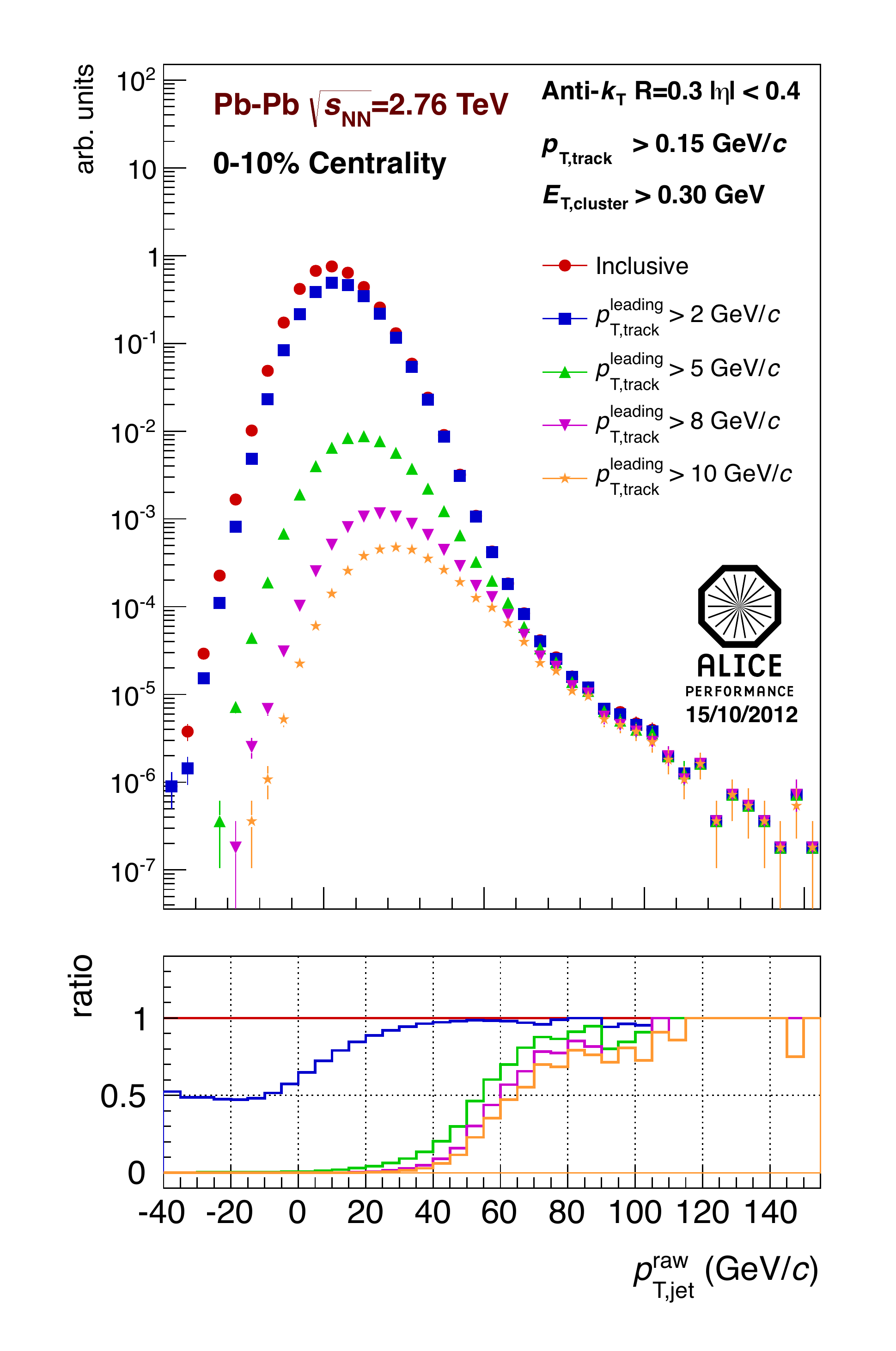}} 
\caption{Raw jet \pT{} spectra in \mbox{Pb--Pb} collisions at $\snn=2.76$~TeV in the 0-10\% centrality range at mid-rapidity:
inclusive is shown in full circles and with 
increasing minimum leading hadron \pT{} requirement in the other symbols (color online). Jets are reconstructed using the \antikt{} algorithm 
with two different resolution parameters, $R=0.2$ (left) and $R=0.3$ (right).}
\label{fig:RawJetSpectraPt}
\end{figure}

The \antikt{}~\cite{antikt-cacciari-08} jet finding algorithm has been employed in its \texttt{FastJet}~\cite{fastjet-cacciari-11} 
implementation. This sequential recombination algorithm has the advantage of
being ``soft-resilent'', namely it is little affected by the soft background~\cite{jet-cacciari-11}. Anti-$k_\mathrm{T}$ jets are pretty regular, 
cone-like shaped around some high-\pT{} particle. 
The uncorrelated energy density from soft processes is large in heavy-ion events and needs to be subtracted 
on an event-by-event basis. 
The main problem in the estimation of the average background is the separation of the UE from the hard scattering.
The method used here has been proposed in Ref.~\cite{bkg-cacciari-08}.
The average background $\rho$ is calculated, event-by-event, as the median of the \pT{}
density (jet \pT{} over jet area) of the \kt{} algorithm~\cite{kt-ellis-93} reconstructed jets.
A jet-by-jet subtraction is performed:
$p_{\rm T,jet}^{\rm raw}=p_{\rm T,jet}^{\rm uncorr}-\rho \times A_{\rm jet}$ ,
where $p_{\rm T,jet}^{\rm uncorr}$ and $A_{\rm jet}$ are respectively the transverse momentum and the area of the jet.
Combinatorial jets, i.e. jets reconstructed out of the soft background, are efficiently
removed by requiring a minimum leading hadron \pT{}. 
Figure~\ref{fig:RawJetSpectraPt} shows the \antikt{} raw jet \pT{} spectra obtained applying different minimum 
leading hadron \pT{} requirement, from 0 (inclusive) to 10~\gevc, for two different jet resolution 
parameters $R=\sqrt{\Delta\eta^2+\Delta\phi^2}$.

\section{Unfolding}
The spectra shown in Fig.~\ref{fig:RawJetSpectraPt} are not directly comparable with model predictions.
In fact the measured values of the observable, the jet \pT{}, are subject to random fluctuations,
due to region-to-region differences in the background momentum density and to the detector response. This means
that each observation is characterized by a true (and unknown) value $p_{\rm T,jet}^{\rm true}$,
and by a measured value $p_{\rm T,jet}^{\rm meas}$. The histograms for $p_{\rm T,jet}^{\rm true}$
and $p_{\rm T,jet}^{\rm meas}$ are related by a convolution through the response matrix 
$\mathrm{RM_{tot}} = \mathrm{RM_{det}} \times \mathrm{RM_{bkg}}$, where $\mathrm{RM_{det}}$ parametrizes
the detector reponse whereas $\mathrm{RM_{bkg}}$ describes background fluctuations.
Unfolding is the numerical procedure that allows to get back the true distribution given the 
measured one and the response matrix $\mathrm{RM_{tot}}$.

\begin{figure}[t]
\includegraphics[width=0.5\textwidth]{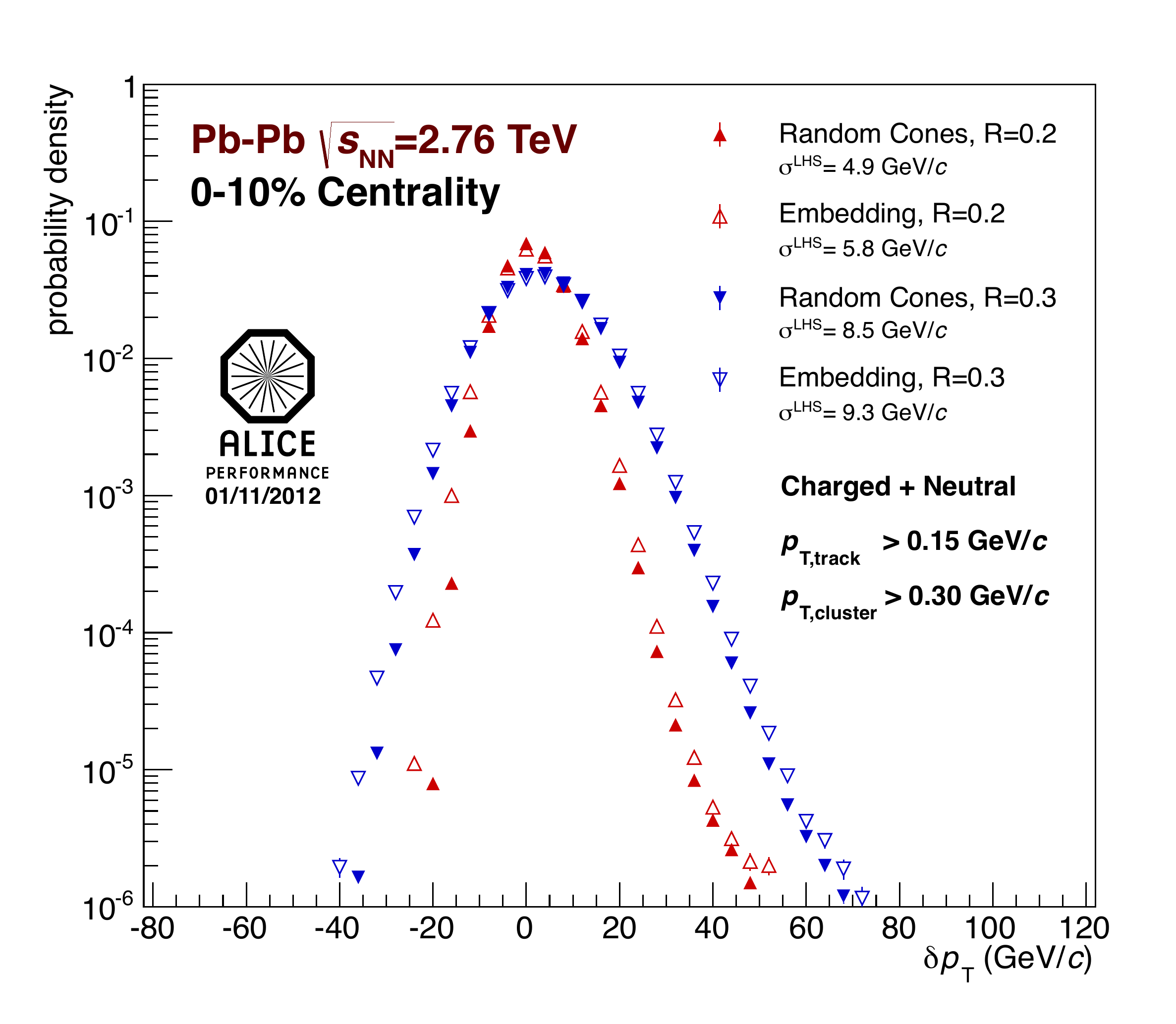}\hspace{2pc}
\begin{minipage}[b]{14pc}\caption{\dpT{} distributions for \mbox{Pb--Pb} collisions at $\snn=2.76$~TeV in the 0-10\% centrality class. 
Two different methods have been implemented: ``random cones'', shown with full symbols, and 
single particle embedding, shown with open symbols (see text for details);
 the distributions for $R=0.2$ are narrower w.r.t.~$R=0.3$ (color online).}
\label{fig:DeltaPtFullV2}
\end{minipage}
\end{figure}
Background fluctuations have been estimated in two different ways~\cite{bkg-alice-12}, namely using random cones
(scalar sum of the \pT{} of all particles found in a cone randomly placed in the event) and 
single particle embedding (with the \antikt{} algorithm).
The residual \pT{} differences due to region-to-region fluctuations are calculated as:
$\delta p_{\rm T} = p_{\rm T,jet} - p_{\rm T,probe} -\rho \times A_{\rm jet}$, 
where $p_{\rm T,probe}$ is the \pT{} of the embedded probe ($p_{\rm T,probe}=0$ for random cones).
The \dpT{} distributions, shown in Fig.~\ref{fig:DeltaPtFullV2}, tell us how much the jet \pT{} 
is smeared due to background fluctuations.

The detector response to jet reconstruction has been studied with pp simulated events, using the PYTHIA6~\cite{pythia6} generator
and the GEANT3~\cite{geant3} transport code.
Jets are reconstructed both at generator level and at detector level. The generator-level and detector-level jets
are matched following a geometrical criterion. 

\section{Results}
This measurement makes use of the unfolding iterative method proposed by D'Agostini~\cite{bayes-dago-95}, which
contains elements of Bayesian statistics.
In Bayesian unfolding the number of iterations plays the role of the regularization parameter. Based on 
the evolution of the covariance matrix and the converging of the solution itself, a number of four iterations
has been chosen as default. 
Indeed after three/four iterations the procedure starts to converge. However, with more iterations
a characteristic fluctuation pattern does appear along with larger variances and (anti-)correlations between
far bins (indicating under-regularization).
Systematic uncertainties have been estimated taking into account variations
of the solution for $\pm 1$ iterations ($\sim 5\%$).


\begin{figure}[t]
\centering
{\includegraphics[width=.44\columnwidth]{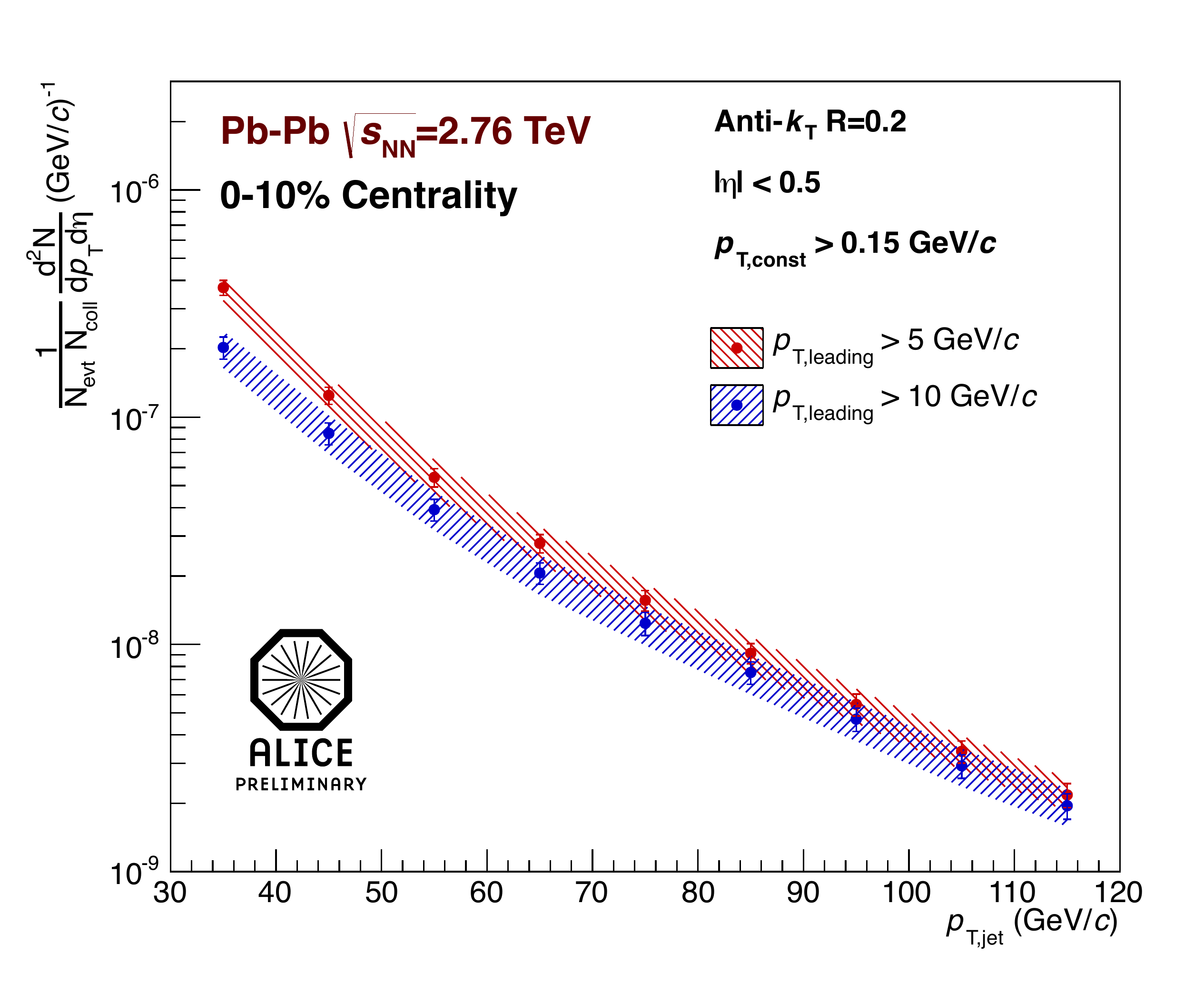} } \quad 
{\includegraphics[width=.44\columnwidth]{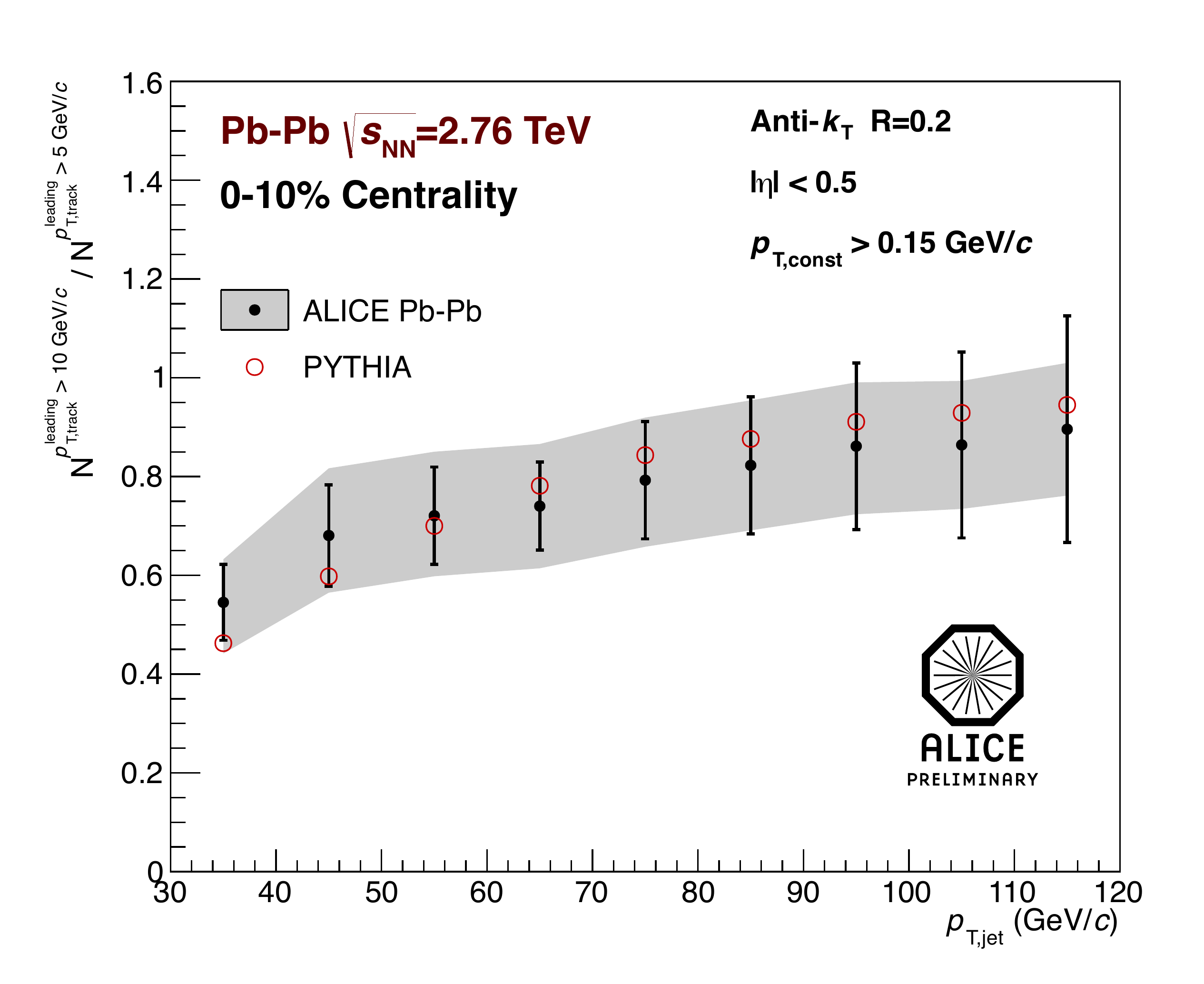} }
\caption{On the left: spectra of reconstructed jets in Pb--Pb collisions at $\snn=2.76$ TeV in the 0-10\% centrality class 
with two different minimum leading hadron \pT{} requirements.  On the right: ratio of the two spectra (solid circles) and comparison
 with a PYTHIA pp expectation (open circles). The bands represent the systematic uncertainty; statistical uncertainties
are shown with bars, when they are larger than markers (colors online).}
\label{fig:FinalSpectraLHRAndRatio}
\end{figure}

To better understand the effect of the leading hadron requirement
in the jet \pT{} spectrum, the analysis was performed for $p_{\rm T, leading} > 5,~10~\gevc$.
Figures~\ref{fig:FinalSpectraLHRAndRatio} show the two spectra with the two different leading hadron thresholds,
and the ratio, compared to a PYTHIA expectation. 

\section{Conclusions}
We have reported on the analysis techniques utilized to reconstruct jets in the 10\% most central Pb--Pb 
collisions at $\snn=2.76$~TeV recorded by ALICE in 2011. The raw jet \pT{} spectra are corrected for the average
background and biased requiring a minimum leading hadron \pT{}. Corrections
for background fluctuations and detector effects are applied via Bayesian unfolding.

The effect of the leading hadron requirement was studied for two different thresholds, 5~\gevc{} and 10~\gevc{}.
The ratio between the two corrected spectra is in reasonable agreement with a PYTHIA simulation, 
which indicates a vacuum-like fragmentation of the jet core.

\section*{References}
\bibliography{biblio}

\providecommand{\newblock}{}
\begin{thebibliography}{10}
\expandafter\ifx\csname url\endcsname\relax
  \def\url#1{{\tt #1}}\fi
\expandafter\ifx\csname urlprefix\endcsname\relax\def\urlprefix{URL }\fi
\providecommand{\eprint}[2][]{\url{#2}}

\bibitem{hadrons-star-03}
Adams J {\em et~al.\/} (STAR Collaboration) 2003 {\em Phys. Rev. Lett.\/} {\bf
  91} 172302

\bibitem{hadrons-phenix-04}
Adler S {\em et~al.\/} (PHENIX Collaboration) 2004 {\em Phys. Rev.\/} C {\bf
  69} 034910

\bibitem{hadrons-alice-10}
Aamodt K {\em et~al.\/} (ALICE Collaboration) 2011 {\em Physics Letters\/} B
  {\bf 696} 30 ISSN 0370-2693

\bibitem{hadrons-cms-12}
Chatrchyan S {\em et~al.\/} (CMS Collaboration) 2012 {\em The European Physical
  Journal\/} C {\bf 72}(3) 1--22 ISSN 1434-6044

\bibitem{dijet-star-06}
Adams J {\em et~al.\/} (STAR Collaboration) 2006 {\em Phys. Rev. Lett.\/} {\bf
  97} 162301

\bibitem{dijet-alice-12}
Aamodt K {\em et~al.\/} (ALICE Collaboration) 2012 {\em Phys. Rev. Lett.\/}
  {\bf 108}(9) 092301

\bibitem{dijet-atlas-10}
Aad G,  {\em et~al.\/} (ATLAS Collaboration) 2010 {\em Phys. Rev. Lett.\/} {\bf
  105}(25) 252303

\bibitem{jet-cms-11}
Chatrchyan S {\em et~al.\/} (CMS Collaboration) 2011 {\em Phys. Rev.\/} C {\bf
  84}(2) 024906

\bibitem{bkg-alice-12}
Abelev B {\em et~al.\/} (ALICE Collaboration) 2012 {\em JHEP\/} {\bf 1203} 053

\bibitem{alice-08}
Aamodt K and other (ALICE Collaboration) 2008 {\em Journal of
  Instrumentation\/} {\bf 3} S08002

\bibitem{ppr-emcal}
Bellwied R (for the ALICE EMCal Collaboration) 2010 {\em Arxiv preprint
  arXiv:1008.0413\/}

\bibitem{antikt-cacciari-08}
Cacciari M, Salam G and Soyez G 2008 {\em JHEP\/} {\bf 04} 063

\bibitem{fastjet-cacciari-11}
Cacciari M, Salam G and Soyez G 2011 {\em Arxiv preprint arXiv:1111.6097\/}

\bibitem{jet-cacciari-11}
Cacciari M, Rojo J, Salam G and Soyez G 2011 {\em The European Physical
  Journal\/} C {\bf 71} 1--21

\bibitem{bkg-cacciari-08}
Cacciari M and Salam G 2008 {\em Physics Letters\/} B {\bf 659} 119--126

\bibitem{kt-ellis-93}
Ellis S~D and Soper D~E 1993 {\em Phys. Rev.\/} D {\bf 48} 3160

\bibitem{pythia6}
Sj{\"o}strand T, Mrenna S and Skands P 2006 {\em Journal of High Energy
  Physics\/} {\bf 5} 1--581

\bibitem{geant3}
Geant3 webpage, http://wwwasd.web.cern.ch/wwwasd/geant/

\bibitem{bayes-dago-95}
D'Agostini G 1995 {\em NIM\/} A {\bf 362} 487

\end{thebibliography}

\end{document}